\documentclass[12pt,preprint]{aastex}

\shorttitle{Nature of Nearby Counterparts to LCBGs}
\shortauthors{C. A. Garland et al.}

\begin{document}

\title{The Nature of Nearby Counterparts to \\ Intermediate
Redshift
Luminous Compact Blue Galaxies \\ I.  Optical/H~{\small{I}}
Properties and  
Dynamical Masses}

\author{C. A. Garland\altaffilmark{1}}
\affil{Institute for Astronomy, University of Hawai`i, 2680
Woodlawn Drive,
    Honolulu, HI 96822}
\email{cagarland@alum.colby.edu}
\altaffiltext{1}{Present address: Natural Sciences Department, 
Castleton State College, Castleton, VT 05735}

\author{D. J. Pisano\altaffilmark{2}}
\affil{CSIRO Australia Telescope National Facility, P.O. Box
76, Epping, NSW
1710, Australia}
\email{DJ.Pisano@csiro.au}
\altaffiltext{2}{Bolton Fellow \& NSF MPS Distinguished International 
Postdoctoral
Research Fellow}

\author{J. P. Williams}
\affil{Institute for Astronomy, University of Hawai`i, 2680
Woodlawn Drive,
Honolulu, HI 96822}
\email{jpw@ifa.hawaii.edu}

\author{R. Guzm\'{a}n}
\affil{Department of Astronomy, University of Florida, 211
Bryant Space Science
Center,
P.O.~Box~112055, Gainesville, FL 32611-2055}
\email{guzman@astro.ufl.edu}

\and

\author{F. J. Castander}
\affil{Institut d'Estudis Espacials De Catalunya/CSIC, Gran
Capit\`a 2-4,
E-08034 Barcelona, Spain}
\email{fjc@ieec.fcr.es}

\begin{abstract}
We present single-dish H~{\small{I}} spectra obtained with
the Green Bank Telescope,
along with optical photometric properties from
the Sloan Digital Sky Survey, of 20 nearby (D~$\lesssim$~70~Mpc)
Luminous 
Compact Blue Galaxies (LCBGs).
These $\sim$L$^\star$, 
blue, high surface brightness, starbursting galaxies were
selected with
the same criteria used to define
LCBGs at higher
redshifts.  We find these galaxies are gas-rich, with
M$_{HI}$ 
ranging from 
5$\times$10$^8$ to  
8$\times$10$^9$~M$_\odot$, and M$_{HI}$~L$_B^{-1}$ ranging from 
0.2 to 2~M$_\odot$~L$_\odot^{-1}$, consistent with a variety of
morphological types
of galaxies.  We find the dynamical masses (measured
within R$_{25}$) span a wide range, from
3$\times$10$^9$ to 1$\times$10$^{11}$~M$_\odot$.
However, at least half have dynamical mass-to-light ratios
smaller than nearby galaxies of all Hubble types,
as found for
LCBGs at intermediate redshifts. 
By comparing line widths and effective radii with local
galaxy populations, we find that LCBGs
are consistent with the dynamical mass
properties of Magellanic (low luminosity) spirals, and the more
massive irregulars and dwarf ellipticals, such as NGC~205.
\end{abstract}


\section{Introduction}

\subsection{Luminous, compact, star forming galaxies in the
distant 
Universe}
The Hubble Space Telescope and advances in ground based
observing have
greatly increased our knowledge of the galaxy population in 
the distant Universe.  However, the nature of these galaxies
and their evolutionary connections to local galaxies remain
poorly understood.  Luminous, compact, 
star forming galaxies appear to represent a prominent phase
in the
early history of galaxy formation \citep{jan03}.  In
particular:  

\begin{itemize}
\item The number density of luminous, compact star forming
galaxies
rises significantly out to z~$\sim$~1 \citep{koo94,
guz97, phi97, lil98, mal99}.  

\item The Lyman Break galaxies at z~$>$~2 seen in the Hubble Deep
Field are characterized by very compact cores and a high
surface brightness
\citep{gia96, low97, wee98}.

\item Sub-millimeter imaging has revealed distant galaxies 
(z~$\sim$~2$-$4), 
 half of them compact objects,
which may be responsible for as much as half of 
the total star formation rate in the early Universe
\citep{sma98}.
\end{itemize}

\noindent
However, little is definitively known of their physical
properties, or how they are related to subsets of the local
galaxy population.

A classification for known examples of 
intermediate redshift (0.4~$\lesssim$~z~$\lesssim$~0.7) luminous, blue, 
compact galaxies, such
as Blue
Nucleated Galaxies, Compact Narrow Emission Line Galaxies,
and
Small Blue Galaxies, has been developed by \cite{jan03} in
order to
be able to choose samples over a wide redshift range.  They
have found
that the bulk of these galaxies, collectively termed Luminous 
Compact Blue Galaxies
(LCBGs), can be distinguished quantitatively from local
normal galaxies
by their blue color, small size, high luminosity,
and high surface brightness. (See \S{2.1} for more detail.)
   
From studies at intermediate redshifts, it has been found
that
LCBGs are a heterogeneous class of 
vigorously starbursting, high metallicity galaxies with an
underlying older
stellar
population \citep{guz96, guz98b}.  While common at
intermediate redshifts, 
they
are rare locally \citep{mal99} and little is known about the
class as a whole,
nor their evolutionary connections to other galaxies.
LCBGs undergo dramatic evolution: At z~$\sim$~1, they are
numerous and have a
total star formation
rate density equal to that of grand-design spirals at that
time.  However,
by z~$\sim$~0, the number density and star formation
rate density of LCBGs has decreased by
at least a factor
of ten \citep{mar94, guz97}.  Since the
LCBG population is morphologically and spectroscopically diverse,
these galaxies are unlikely 
to evolve into one homogeneous galaxy class.  \cite{koo94}
and \cite{guz96}  
suggest that a subset of LCBGs at intermediate redshifts
may be the progenitors of local low-mass dwarf elliptical 
galaxies such as NGC~205.  Alternatively,
\cite{phi97} and \cite{ham01} suggest that others
may be disk galaxies in the process of building a 
bulge to
become local L$^\star$ spiral galaxies.  

Clearly, to determine the most likely
evolutionary scenarios for
intermediate redshift LCBGs, it is necessary to know their
masses and the
timescale of their starburst activity.
Are they comparable to today's massive or low-mass galaxies?
Are they small starbursting galaxies which will soon exhaust their
gas and eventually fade?  Or
are they larger galaxies with only moderate amounts of star
formation?
Only kinematic line widths that truly reflect the masses of
these galaxies, as well as measures of their gas content and
star formation rates, can answer these questions. 
Using ionized gas emission line widths, \cite{koo94},
\cite{guz96}, and
\cite{phi97},
have found that LCBGs have mass-to-light
ratios approximately ten times smaller than typical local L$^\star$
galaxies.
 However, since ionized gas emission lines may originate primarily from
the central regions of galaxies,
their line widths
may underestimate the gravitational potential
\citep{bar01b}. 
 H~{\small{I}} 
emission lines provide a
better estimate of the total galaxy
mass as they measure the gravitational potential
out to larger galactic radii.
Observations of
both H~{\small{I}} and CO (the best tracer of cold
H$_2$), combined
with star formation rates, are necessary to estimate the
starburst
timescales.

\subsection{A local sample of LCBGs}

With current radio instrumentation, H~{\small{I}} and CO 
can only easily be measured in very
nearby LCBGs, at distances $\lesssim$~150 Mpc for H~{\small{I}},
and $\lesssim$~70 Mpc for CO.
Therefore, to understand the nature and 
evolutionary possibilities of
higher redshift
LCBGs, 
we have undertaken a survey in H~{\small{I}} 21~cm 
emission and multiple rotational
transitions of CO of a 
sample of 20 local LCBGs, drawn from the Sloan
Digital Sky Survey \citep{yor00}.  
This work, Paper I, reports the optical photometric
properties of our sample and 
the results of the H~{\small{I}} 21~cm portion of the
survey, including
dynamical masses and comparisons with
local galaxy types.
Paper II \citep{gar04}
will report the results of a survey of the molecular gas
conditions.
Knowledge of the dynamical masses, combined with 
gas masses and star formation rates, 
constrains the evolutionary possibilities of these galaxies.

Nearby blue compact galaxies (BCGs) have been
studied extensively at 
radio and optical wavelengths since 
\cite{zwi64} originated the term ``compact galaxy'' and
\cite{sar70} distinguished between ``red'' and ``blue''
compact galaxies. 
The term BCG typically refers to galaxies with a compact
nature,
a high mean surface brightness, and emission lines
superposed on a blue
continuum.  
However, many different selection criteria have been used,
leading to 
various definitions of BCGs and samples with a range of
properties. 
For example, 
the term ``dwarf''  has been used to mean BCGs
fainter than $-$17 (e.g. Thuan \&
Martin 1981; Kong \& Cheng 2002) or $-$18 blue magnitudes
(e.g. Taylor et al. 1994),
or an 
optical diameter less
than 10~kpc \citep{cam93}.   The term ``blue'' has
been used to mean blue on the Palomar Sky Survey Plate (e.g.
Gordon \&
Gottesman 1981), or to have emission lines superposed on a
blue
background (e.g. Thuan \& Martin 1981).  The term
``compact'' has
been used to mean smaller than 1 kpc in optical diameter 
(e.g. Thuan \& Martin 1981), or qualitatively compact (e.g.
Doublier 
et al. 1997).  Some, for example \cite{thu81}, began using the term
Blue Compact
Dwarf to refer to the common low luminosity, low metallicity
BCGs.  As high
luminosity ($\sim$L$^\star$), nearby BCGs are rare, many
began to use the
term Blue Compact Dwarf for all nearby BCGs, regardless of
luminosity.

There are only a few of the rare local LCBGs
(by the \cite{jan03} definition) in previous BCG surveys.
Note that \cite{ber02} have recently been studying local
luminous BCGs.  However, their selection criteria 
are not as stringent as that of
\cite{jan03}, and they have been focusing on 
very low metallicity (less than
15\% solar) galaxies.
Intermediate redshift LCBGs 
with rest-frame properties matching the class definition
of \cite{jan03}
have
metallicities at least 40\%~solar \citep{guz96}.
Four of our local LCBGs have metallicities available in the
literature,
which range from 40~$-$~70~\%~solar \citep{hun86, cal97}.  
By using the selection criteria of \cite{jan03}, we
ensure that our study is of those local LCBGs
defined to be analogs to the widely studied higher redshift
LCBGs.

The observations, including sample selection and data
reduction, are 
described in $\S$2.  The optical photometric properties, 
H~{\small{I}} 21~cm spectra, measurements, and derived
properties
are presented in $\S$3, and analyzed in $\S$4.  We compare
the derived physical properties to local normal galaxies and
higher redshift LCBGs in $\S$5, and conclude in $\S$6.
We assume H$_0$~=~70~km~s$^{-1}$~Mpc$^{-1}$ throughout. 
When we compare
our results to those of other authors, we scale their
results to 
this value.

\section{The Observations}

\subsection{Sample Selection}
\subsubsection{Selection Criteria}
When \cite{jan03} compared intermediate redshift
LCBGs with local normal galaxies, they
found that LCBGs can be isolated quantitatively on the basis
of color, surface brightness, image concentration, and
asymmetry.
Color and surface brightness were found to give the best
leverage for 
separating LCBGs from normal galaxies.  Specifically, LCBGs
can be defined by a region limited to
B$-$V~$<$~0.6, and 
a B-band surface brightness within the half-light radius, 
SBe, brighter than 21~B-mag~arc~sec$^{-2}$.  This simple
definition differs only slightly from the formal definition
in \cite{jan03} which uses a color-dependent SBe.
\cite{jan03} also 
applied a luminosity cut-off of 
25\%~L$^\star~($M$_B$~$<$~$-$18.5), 
to distinguish LCBGs from blue
compact galaxies.  LCBGs are not so extreme that they are
completely
separated from the continuum of normal galaxies.  The
sharp borders
used to classify them are artificial, but serve the purpose
of 
defining similar objects over a range of redshifts.  
LCBGs at intermediate redshifts (z $\sim$ 0.6) can be
studied in deep spectroscopic surveys (I $\sim$ 24),
while the brightest LCBGs at high redshifts (z $\sim$ 3) require very 
deep spectroscopic
surveys (I $\sim$ 26).  Therefore, LCBGs selected in this manner are 
observable over
a wide range of redshifts.  

Using these color, surface brightness, and luminosity
criteria, 
we selected our 
sample of local LCBGs 
from the Sloan Digital Sky Survey (SDSS). 
Begun in 2000, this survey will ultimately
image one quarter of the sky  using a large format
CCD camera on a 2.5~m telescope at Apache Point Observatory
in New
Mexico.  Images are taken in five broad bands (u, g, r, i,
z) which
range from 3540~\AA~(u) to 9130~\AA~(z).  The survey has a
limiting
magnitude of
22.2 in g~(4770~\AA) and r~(6230~\AA), 
our two bands of interest.
After searching 
through approximately one million galaxies ($\sim$1500 degree$^2$ on the
sky), we identified 
only 16 nearby (D~$\lesssim$~70~Mpc) LCBGs.  This distance
cut-off was chosen to ensure our galaxies could be detected
quickly in 
both H~{\small{I}} and
CO. 
We added four
Markarian galaxies from the literature which fulfilled our
selection
criteria and were not yet surveyed by SDSS, 
for a total of 20 local LCBGs.  The color and
surface brightness
characteristics of our local sample, as well as higher
redshift (0.4~$\lesssim$~z~$\lesssim$~1)
samples of
LCBGs, are compared to other nearby galaxies in Figure 1.
We note that while our sample is not complete, it is representative
of local LCBGs (see Castander et al. 2004).

\subsubsection{Galaxy Properties}
As at higher redshifts
\citep{phi97}, our local LCBG sample is
a morphologically heterogeneous mixture of galaxies, including Sb to Sc
spirals, 
S0 galaxies, polar ring galaxies, peculiar galaxies, and 
H {\small{II}} liners and starbursts,
as classified by
HyperLeda\footnote{http://www-obs.univ-lyon1.fr/hypercat/}
 and the NASA$/$IPAC
Extragalactic
Database\footnote{http://nedwww.ipac.caltech.edu/} (NED).
Approximately one quarter of these galaxies are not 
classified in either database.  
Roughly one third are members of multiple, sometimes interacting or
merging, systems. 
SDSS and
Digitized Sky Survey
images of our sample of 20 galaxies are shown in
Figures 2 and 3.  
The properties of our sample of local LCBGs are described below,
and listed in Table 1.

\emph{Source.} Full SDSS galaxy designation of the form SDSS 
JHHMMSS.ss$+$DDMMSS.s, in the J2000 system.
In the remainder of the paper,
individual galaxies are referred to by an abbreviated SDSS
name of the
form SDSSJHHMM$+$DDMM.  The four non-SDSS galaxies are
referred to by their
Markarian (Mrk) names. 

\emph{Alternate Name.} Mrk and/or NGC designation, if any.

\emph{D$_{OPT}$.} Hubble distance, in Mpc,
 calculated from the optical redshift 
(from SDSS, except for the non-SDSS galaxies for
which we use NED redshifts.)

\emph{M$_B$.} Absolute blue magnitude calculated from the 
apparent blue magnitude, m$_B$, 
using D$_{OPT}$.
For the SDSS galaxies, m$_B$~=~g~+~0.30(g$-$r)~+~0.18\footnote{
The relationships for m$_B$, B$-$V, and r$_e$(B) are based
on synthetic spectra which fit the observed spectral energy
distribution of LCBGs.}.
The
r and g magnitudes were calculated from the SDSS
``petrosian'' and
``model'' magnitudes:  r~=~r(petro) and 
g~=~r(petro)~$-$~[r(model)~$-$~g(model)] \citep{sto02}.
The SDSS Galactic reddening corrections were applied, but
no K corrections were applied
as our galaxies are nearby, and many are of unknown
spectral
type.  The K correction is at most 0.015 magnitudes in
r and 0.038 magnitudes in g \citep{pog97}.  No correction
was applied
for extinction due to inclination, also because of the
uncertainty in
spectral types.  For the non-SDSS galaxies, the 
``total apparent blue magnitudes'' were used from HyperLeda, and
corrected for
Galactic reddening using extinction values from NED.

\emph{B$-$V.} The color transformation is
B$-$V~$=$~0.900(g$-$r)~+~0.145$^5$, where
the magnitudes were calculated as above, for the SDSS
galaxies.
The ``total B$-$V colors'' from HyperLeda were
used for the non-SDSS galaxies.  They were corrected
for reddening using extinction values from NED.

\emph{SBe.} Average surface brightness in the B band
within the half light or effective radius in 
B-mag~arc~sec$^{-2}$. 
For the SDSS galaxies,
SBe(B)~=~m$_B$~+~2.5~$\log$~[2$\pi$r$_e^2$(B)], where
r$_e$(B),
the effective or half-light radius in the B band, 
was calculated from the ``petrosian'' 
effective radii in the g and r bands 
[r$_e$(B)~=~1.30~$r_e$(g)~$-$~0.300~$r_e$(r)]$^5$.
We 
corrected SBe for cosmological dimming by subtracting 
7.5~$\log$(1 + z), 
where the redshift, z, was calculated using the H~{\small{I}} velocity
($\S$3.1).
For the non-SDSS galaxies, 
we used SBe from HyperLeda (termed the
``mean effective surface brightness''),
and then
corrected it for cosmological dimming in the same way.

\emph{NED Type.} Morphological type as given by NED,
which does not use a homogeneous system.

\emph{HyperLeda Type.} Morphological type as given by HyperLeda, which
uses the
\cite{dev91} system.

\emph{Other Sources in Beam?} Indicates whether or not 
each galaxy has other sources at similar velocities
within the 9$^\prime$.2 beam of the Green Bank Telescope (used for
the H~{\small{I}} observations), as given by
NED.

\subsubsection{Comparison of SDSS and HyperLeda properties}
We compared HyperLeda and SDSS magnitudes for our SDSS selected galaxies
to ensure that there are no systematic offsets between the two.  Fourteen 
galaxies had magnitudes measured in both systems.
The median difference
between SDSS and HyperLeda magnitudes (as calculated in
$\S$2.1.2) is 0.1~magnitudes.  Since the standard deviation of these
differences is $\pm$~0.8~magnitudes,
we estimate the uncertainty in the median magnitude difference
to be 0.2~magnitudes.  Therefore, there is no significant difference
between the HyperLeda and SDSS magnitudes.
Unfortunately, while fourteen galaxies have magnitudes measured
in both systems, 
only two SDSS selected galaxies have B$-$V and SBe entries in HyperLeda; 
for these two galaxies there are no significant differences
between the HyperLeda and SDSS values.

\subsection{Observations}

H~{\small{I}} 21~cm 
observations of 19 of the 20 nearby LCBGs in our sample
were made with the
100~meter Green
Bank Telescope (GBT) at the National 
Radio Astronomy Observatory\footnote{
The National Radio Astronomy Observatory is a facility of the 
National Science Foundation operated under cooperative agreement 
by Associated Universities, Inc.}
in Green Bank, West Virginia
between 2002~November~30 and December~6.  
The H~{\small{I}} spectrum of SDSSJ0934$+$0014 was acquired
by 
\cite{fis03}
during GBT commissioning on 2001~November~29. 
The main beam half power width is 9$^\prime$.2 at 21~cm \citep
{loc98}.  Both
linear and circular polarizations were observed using the
L-band receiver 
(1.15~$-$~1.73~GHz).  Position-switching was
used with an offset of
$-$18$^\prime$ in Right Ascension.  
The spectral processor was used with a
bandwidth of 20~MHz ($\sim$4000 km~s$^{-1}$) for most galaxies,
although a few were
observed with a bandwidth of 5~MHz ($\sim$1000 km~s$^{-1}$).  
The sample time was 60~s,
with each
galaxy observed for between 6 and 30 minutes, for a peak signal-to-noise
of at least five.  The only exceptions were SDSSJ0218$-$0757 and
SDSSJ0222$-$0830, marginal detections, which were observed for
52 minutes each.

\subsection{Reduction}

The initial data calibration and reduction was performed
using the AIPS++
single dish analysis environment ``dish.''  The
position-switched data were
calibrated in the standard way taking the difference of the
on and off
scans divided by the off scans.  The dish package
automatically calibrates the data into temperature units
using tabulated values for the
position of the
telescope.  The individual scans were then combined and a
first order
baseline was fit to the line-free regions and removed.  Our
wide bandwidths ensured a sufficient line-free region.
The
individual
polarizations were then combined, except when one
polarization was
significantly noisier than the other, or was affected
by radio interference.  Finally, each
spectrum was
smoothed to $\sim$12~km~s$^{-1}$ channels using
boxcar smoothing.  

The rest of the reduction and analysis was done
using our own
procedures written in Interactive Data Language (IDL).     
To convert our data from temperature to flux density units, 
we observed the radio galaxy 3C295.  It makes
an ideal calibration 
source as it has not varied by more than 
$\sim$1$\%$ since 1976 for 2.8~cm~$\leq$~$\lambda$~$\leq$~21~cm
\citep{ott94}.  Comparing our observations of 3C295 with
those of
\cite{ott94}, we found the gain of the GBT 
to be 1.9~K~Jy$^{-1}$ and applied this calibration to our data.
Finally,
obvious 
noise spikes were
clipped out of the data.
Data of SDSSJ0934$+$0014, the galaxy observed during
telescope commissioning, 
were
reduced and calibrated by \cite{fis03}.  The spectrum
was then smoothed to 
$\sim$12~km~s$^{-1}$ channels.

The central 800~km~s$^{-1}$ of the final
H~{\small{I}} 21~cm spectra of the 20 local LCBGs are shown
in Figure 4.  Each galaxy's velocity calculated from optical
redshifts is indicated with a triangle.  Dashed lines indicate
the 20\% crossings used to measure the line widths ($\S$3.1).

\section{Results}

All 20 galaxies were detected in the 21~cm line of H
{\small{I}}, although
two, SDSSJ0218$-$0757 and SDSSJ0222$-$0830, were only
detected at the 3 and 4 $\sigma$ (respectively) level.
H~{\small{I}} measurements
and derived optical and H~{\small{I}} quantities 
are listed in Tables 2, 3 and 4 and described in the following
sections.
Since the GBT beam is 9$^\prime$.2 at 21~cm, the emission
from
the target galaxies and any other galaxies within the beam and
at similar velocities
are blended into one
spectrum. (See Table 1 for those galaxies with other sources
at similar velocities within the GBT beam.)

We find fairly broad H~{\small{I}} profiles 
(126~km~s$^{-1}$~$<$~W$_{20}$~$<$~362~km~s$^{-1}$)
and a variety of
profile
shapes, including double-horned, flat-topped and Gaussian.  
Over a third
of our profiles appear asymmetric 
and only half of these are from galaxies
known to have other sources within the GBT beam.
These asymmetries may indicate asymmetries in the gas
density distribution and/or disk kinematics, or
unidentified galaxies within the beam
\citep{ric94}. 

\subsection{Dynamical Masses}
Estimates of the dynamical masses of our local sample of
LCBGs constrain
 their evolutionary possibilities.  
We estimate the dynamical
mass, M$_{DYN}$,
within a radius, R, as
\begin{equation}
M_{DYN}~(< R) =~~c_2~~\frac {V^2_{ROT} ~R} {G}  ~~~.
\end{equation}
The constant c$_2$ is a geometry-dependent factor which depends on the galactic
light profile.  For example, the King models of \cite{ben92} give
0.9~$\lesssim$~c$_2$~$\lesssim$~1.5, depending on the ratio of tidal to 
core radius.
As we do not have information on the light profiles of our galaxy sample, 
we simply choose c$_2$ = 1.

The line width at 20$\%$, corrected for the effects
of inclination, W$^i_{20}$, is 
used as a measure of
twice the maximum rotational velocity, V$_{ROT}$ \citep[e.g.][]{tul85}.  
W$_{20}$ was measured at the points equal to
20\% of the peak flux.
The first crossing at 20$\%$ on each side of the emission
line was used, once the spectrum
became distinguishable from the noise.
These crossings are indicated by dashed lines in Figure 4.
We find W$_{20}$ ranging from 
126~$-$~362~km~s$^{-1}$.  When \cite{phi97} studied LCBGs at
intermediate redshifts, they found a similar range of line widths, using 
optical emission lines.  The recessional velocities in the barycentric
system, V$_{BARY}$,
 were calculated as the midpoint between
the 20$\%$ crossings.  The uncertainties listed in Table 2 for W$_{20}$,
V$_{BARY}$, and the galaxy distance (D) are from uncertainties in 
measuring the exact W$_{20}$ crossings.

Two galaxies have multiple crossings at the 20\% flux level, once
the spectrum is distinguishable from the noise.  
One of these,
SDSSJ0834$+$0139, has a wing on the H~{\small{I}} spectrum which extends
over 200~km~s$^{-1}$.
Neither SDSSJ0834$+$0139 nor its nearby companion have optical velocities
coincident with the peak of the H~{\small{I}} spectrum; instead, both
have velocities in the wing of the spectrum (see Figure 4).  For this reason, 
we 
used W$_{20}$ measured from the last crossing, which is at
the edge of the wing.  
The other galaxy with multiple crossings at 20\% is SDSSJ0904$+$5136, which 
has a low
level extension to one side of the spectrum.  We used the first crossing at 
20\%.
The uncertainties in these two W$_{20}$ measurements are 
reflected in the large errors associated
with W$_{20}$ and derived quantities.
We have initiated an observing program at the Very Large Array (VLA) to
map these local LCBGs in H~{\small{I}} to disentangle
target galaxy emission from any other galaxies in the field, improving
the mass estimates.

To correct the line width for the effects of inclination (\emph{i}),
we divided W$_{20}$ by $\sin$(\emph{i}), to give
W$^i_{20}$.  Each SDSS galaxy's inclination was approximated
from the SDSS data by:
\begin{equation}
{i} = \arccos \frac {minor~axis} {major~axis} ~~.
\end{equation}
The SDSS isophotal major and minor axes in the g-band~(4770~\AA)
were used, except
for SDSSJ0943$-$0215
where no g band data were available.  In that case,
r-band~(6230~\AA) data were used.  
The SDSS isophotal axes are derived from the ellipticity of
the 25~magnitude~arc~sec$^{-2}$ isophote in each band.
Inclinations derived
from SDSS isophotal axes in r, i~(7630~\AA), and z~(9130~\AA)
 bands agree with inclinations derived
from g-band data to within 8$^\circ$.  
The inclinations from
HyperLeda
were used for the Markarian galaxies.
Thirteen of our SDSS galaxies also have inclinations
available in HyperLeda, which are 
calculated from the apparent flattening and
morphological type of the galaxies.  We estimate the dispersion of the 
differences
between SDSS and HyperLeda inclinations is 12$^\circ$. 

A further correction can be made to W$^i_{20}$ to account
for random motions, giving W$^i_R$.  As outlined in \cite{tul85}, 
\begin{equation}
W_R^2 = W_{20}^2 + W_t^2 - 2W_{20}W_t[1 - \exp-(W_{20} / W_c)^2] - 2W_t^2\exp-(W_{20}/W_c)^2
\end{equation}
and
\begin{equation}
W_R^i = \frac {W_R} {\sin(i)} ~~.
\end{equation}
W$_t$ is the random motion component of the line width; 
W$_t$~=~38~km~s$^{-1}$.
W$_c$ characterizes the transition region
between linear and quadrature summation of rotational and dispersive
terms; W$_c$~=~120~km~s$^{-1}$. The formula degenerates to linear summation 
for giant galaxies,
and quadrature summation for dwarf galaxies \citep{tul85}.
This correction for random motions
decreases the line width of the local LCBG sample by 26~to~38~km~s$^{-1}$, 
depending on the rotational velocity.

It is crucial, when comparing dynamical masses from different studies, that
the rotational velocities be calculated in the same way.  In general, we 
calculate the 
rotational velocities
from W$_R^i$.  However, when we wish to compare our sample with others
\citep[e.g.][]{rob94} who have not applied this correction for 
random motions to their line widths, we also do not apply this correction.
We indicate dynamical masses calculated
without correcting W$^i_{20}$ for random motions as M$^{NR}_{DYN}$.  Unless
indicated as such, all dynamical masses are calculated using the line width
corrected for random motions, W$_R^i$.  Note that in all cases, the line widths
have been corrected for inclination.

It is also crucial, when comparing
dynamical
masses from different studies, that they be measured within
the same
radius. Typically, the  
H~{\small{I}} radius (R$_{HI}$) is measured 
at 1~M$_\odot$~pc$^{-2}$ and is used to estimate
the total enclosed mass of a galaxy.  However, it is only possible to measure
R$_{HI}$ with interferometers in nearby galaxies.  By practical
necessity then,
some other radius must be adopted for our sample and the galaxy's mass
is assumed to be spherically distributed in that radius.  
Two different radii are commonly used:   
R$_e$, the effective or half-light radius, and 
R$_{25}$, the isophotal radius at the limiting surface
brightness of 
25~B-magnitudes~arc~sec$^{-2}$.   

We have measurements of both R$_e$ (in r and B band from
SDSS) and R$_{25}$ (in B-band from HyperLeda) for most galaxies.
\cite{vit96a, vit96b} found that
R$_{25}$~=~2.5~R$_e$ for nearby emission line galaxies measured in
r-band.
When we compare our values of R$_e$ to R$_{25}$,
we find medians of 
\begin{equation}
\langle \frac{R_{25}} {R_e(r)} \rangle = 4.1 \pm 0.7
\end{equation}
and 
\begin{equation}
\langle \frac{R_{25}} {R_e(B)} \rangle = 4.0 \pm 0.9~~.
\end{equation}
For the galaxies with no R$_{25}$ 
(SDSSJ0218$-$0757,
SDSSJ0222$-$0830 and SDSSJ1118$+$6316) or R$_e$ (non-SDSS galaxies) 
measurements
available, we estimate these quantities from the above relations.

We find our local sample of LCBGs have dynamical masses,
measured
within R$_{25}$, ranging from 
3$\times$10$^9$ to 1$\times$10$^{11}$~M$_\odot$, with a
median of 
3$\times$10$^{10}$~M$_\odot$.  
The
dynamical masses measured within the effective radius 
(measured in both the r and B-band)
range from 8$\times$10$^8$ to 3$\times$10$^{10}$~M$_\odot$,
with a median of 
8$\times$10$^{9}$~M$_\odot$.

Figure 5 compares the dynamical masses measured for local
LCBGs to nearby galaxies of Hubble type S0a through Im, where
``m'' indicates Magellanic, or low luminosity \citep{rob94}.  Note that 
\cite{rob94} did not correct line widths for random motions as we did.
Therefore, in Figure 5 we compare 
M$^{NR}_{DYN}$(R~$<~$R$_{25}$) which have not been
corrected for random motions.  These dynamical masses are higher
than those we list in Table 3, which have been corrected for random motions.
As seen in Figure 5, some local LCBGs have dynamical masses as large as
L$^\star$ galaxies ($\sim$10$^{11}$~M$_\odot$, Roberts \& Haynes 1994).  
However, 
at least 75$\%$ have dynamical 
masses approximately an order of magnitude smaller than typical
local L$^\star$ galaxies, consistent with observations of LCBGs at
higher redshifts (as discussed in $\S$ 1.1).
However,
our sample includes galaxies down to
0.25~L$^\star$.  Galaxies of this luminosity typically
have
dynamical masses (within R$_{25}$)
$\sim$2$\times$10$^{10}$~M$_\odot$ \citep{rob94}.  
Therefore, it is more appropriate to compare
mass-to-light ratios; See $\S$4.1.

We estimate the random errors associated with the dynamical masses to be
approximately 50\%.  This estimate is made from uncertainties in the inclination
($\pm$~12$^\circ$, from the earlier
comparison of HyperLeda and SDSS inclinations);
uncertainties in radii measurements (as given by HyperLeda and SDSS); 
and uncertainties in measuring W$_{20}$ (as discussed above).
However, the random errors are overwhelmed by the systematic uncertainties 
when 
estimating dynamical masses.  We have chosen a structural constant of c$_2$~=~1, but
this could be as high as 1.5, as discussed earlier.  Even more significant is that
45\% of our galaxy sample have other galaxies at similar velocities
in the beam of the GBT.  As the H~{\small{I}} emission from all sources in the beam
and within the bandwidth
is blended into one spectrum, we may be overestimating the masses of these galaxies.
Finally, a fifth of our galaxies have low inclinations 
($<$~40$^\circ$) 
which lead to large
corrections and therefore increasing uncertainties in the
rotational velocities and overestimations of the dynamical masses.  The dynamical
masses we report may be viewed as upper limits~--~we are certainly overestimating the
masses in many cases, but not underestimating them.

We are pursuing a follow-up program with both the Arecibo Radio Telescope and the VLA
to address these issues.  At 21~cm the Arecibo Radio Telescope has
a beam size of 3$^\prime$.1~$\times$~3$^\prime$.5, 
a third the size of the GBT beam.  This decreases the
number of target galaxies observed with other galaxies in the beam, decreasing
the number of galaxies with overestimates of dynamical masses.  We have already
observed over 40 LCBGs at Arecibo, but results are not yet available.  The 
VLA in ``B'' configuration
provides the ability to map our galaxies in 21~cm emission to a resolution
of 5$^\prime$$^\prime$.  We have begun a survey of those LCBGs with companions.  VLA
maps allow us to disentangle the emission from each galaxy and more accurately
estimate the dynamical mass of the target galaxy.

\subsection{H~{\small{I}} Content}
Along with dynamical mass, a knowledge of the amount of fuel, i.e. atomic and
molecular
gas, available for the starburst
activity, is critical for narrowing down the evolutionary
possibilities for LCBGs.
When combined with star formation rates, this
gives an estimate
of the maximum length of the starburst at the current rate of star formation.  
We defer a full discussion
of this until
Paper II \citep{gar04} where we present our measurements of the CO content
of these
galaxies, but present the H~{\small{I}} results here.

The H~{\small{I}} masses are given by:
\begin{equation}
\frac {M_{HI}} {M_\odot} = 2.356 \times 10^5 ~ \frac
{D_{HI}^2} {Mpc^2} 
~ \int 
\frac {S~dv} {Jy~km~s^{-1}}
\end{equation}
\citep{rob94},
where the 
distance (measured from H~{\small{I}}) is D~=~V$_{BARY}$~H$_o^{-1}$.
The total H~{\small{I}} flux, $\int~{S~dv}$,
was calculated by numerically integrating
under the spectrum where it is distinguishable from the
noise.  For the
two marginal detections, SDSSJ0218$-$0757 and
SDSSJ0222$-$0830, 
H~{\small{I}} masses were calculated using the optical
(SDSS) recessional velocities.

We find
local LCBGs have H {\small{I}} masses ranging 
from  5$\times$10$^8$ to 
8$\times$10$^9$~M$_\odot$, with a median of
5$\times$10$^9$~M$_\odot$.  These span the range
of H~{\small{I}} masses in nearby galaxies across the Hubble
sequence; the
median is that of nearby
late-type spiral galaxies \citep{rob94}.
The uncertainties listed for M$_{HI}$ (Table 3)
include uncertainties
in the distance and integrated flux density, as listed in
Table 2.  However, as in the discussion of dynamical mass
uncertainties in $\S$3.1, the H~{\small{I}} masses for nearly
half of our galaxies are most likely overestimates due to the
presence of other galaxies within the GBT beam.

\section{Analysis}

\subsection{Mass-to-Light Ratios}
Dynamical mass-to-light ratios indicate whether our
local sample of LCBGs are
under-massive for their luminosities.
We calculated the total blue luminosities, L$_B$, of our
galaxies
from M$_B$, assuming M$_{B\odot}$~=~5.48 \citep{rob94}.
We find the median mass-to-light ratio, 
M$_{DYN}$(R $<$ R$_{25}$)~L$_B^{-1}$~=~3~M$_\odot$
L$_{B\odot}^{-1}$, with a minimum
of  0.6 and a maximum of  9~M$_\odot$~L$_{B\odot}^{-1}$.  

\cite{rob94} find that 
M$_{DYN}$(R $<$ R$_{25}$)~L$_B^{-1}$ ranges from 
3.6~$-$~10~M$_\odot$~L$_{B\odot}^{-1}$ in local galaxies across the
Hubble sequence.  The median
M$_{DYN}$(R $<$ R$_{25}$) L$_B^{-1}$ is fairly constant 
ranging from
5 to 7~M$_\odot$~L$_{B\odot}^{-1}$.
However, as discussed in $\S$3.1, to compare
our mass-to-light ratios to those of local, normal Hubble-types 
in \cite{rob94}, we must use dynamical masses which have not
been corrected for random motions.  
Figure 6 compares such mass-to-light ratios for our
local sample of LCBGs and nearby Hubble types.  While some
local LCBGs have mass-to-light ratios equal to or greater than
the median for local Hubble types, approximately half the
local LCBGs have smaller mass-to-light ratios.  Many of the
galaxies with mass-to-light ratios typical of local Hubble types have
other galaxies within the beam, possibly leading to an overestimation
of the dynamical mass-to-light ratio of the target galaxy.  
Therefore, in general, LCBGs tend to have mass-to-light ratios
smaller than local normal galaxies.  That is, most are small
galaxies undergoing large amounts of star formation and not simply
large galaxies with moderate amounts of star formation.
 
\subsection{How gas-rich are LCBGs?}

We calculated the gas mass fraction, 
M$_{HI}$~M$_{DYN}^{-1}$~(R~$<$~R$_{HI}$) for the local sample
of LCBGs as a measure of their atomic gas richness.  We do not have
interferometric observations of these galaxies, so we
estimated the hydrogen radius,
R$_{HI}$, from optical radii.  
\cite{bro94} found for nearby spiral galaxies that 
R$_{HI}$~=~2~R$_{25}$, where R$_{HI}$ is measured
at the 1~M$_\odot$~pc$^{-2}$ level.  
However, \cite{van98} found R$_{HI}$ ranging from 
 3~$-$~5~R$_{25}$
for a small sample of local H {\small{II}} galaxies which
are similar
to LCBGs, but much less luminous (M$_B$~$\gtrsim$~$-$16).
\cite{mar98} studied a large sample of nearby galaxies
and found that galaxies with smaller optical radii
have larger H~{\small{I}} extensions.
\cite{tay94} have measured R$_{HI}$ in one of our LCBGs, Mrk 314,
using the VLA. They find R$_{HI}$~=~4~R$_{25}$.
Therefore,
although we estimated   
R$_{HI}$ as 2~$\times$~R$_{25}$ following \cite{bro94}, we note it may be an 
underestimate, which would cause us to overestimate the gas mass fraction.
We are undertaking interferometric observations of some of our
local sample of LCBGs to directly measure R$_{HI}$.  This will be
addressed in a future paper.  

In local
LCBGs, we find the gas mass fraction, M$_{HI}$~M$_{DYN}^{-1}$,
 ranges from  0.03 to
 0.2,
with a median of  0.08.  
That is, it ranges
from normal Hubble type spirals ($<$~0.1 \citep{rob94}) to
very gas
rich galaxies such as nearby H~{\small{II}} and irregular
galaxies studied by
\cite{van98} with gas fractions ranging from   
0.2~$-$~0.4.
This is similar to the range of  0.01~$-$~0.5 found by
\cite{pis01}
for local blue compact galaxies, most of them less luminous
than LCBGs.

The fraction of hydrogen mass to blue luminosity provides a
distance
independent alternative measure
of the gas richness of galaxies.
We find our local LCBGs have M$_{HI}$~L$_B^{-1}$ 
ranging from  0.09 to  2~M$_\odot$~L$^{-1}_{B\odot}$, 
with a median of 0.4~M$_\odot$~L$^{-1}_{B\odot}$.  This spans
the
range of nearby galaxies from S0a through Im, 
the median corresponding to late-type spirals \citep{rob94,
bet03}.  
\cite{gor81}
studied a range of blue compact galaxies, ranging from faint 
blue compact dwarfs to LCBGs,
and found the same median value, 0.4~M$_\odot$~L$^{-1}_{B\odot}$.

\subsection{The Tully-Fisher relationship and local LCBGs}
The Tully-Fisher relationship, a relation between line width
and luminosity,
is a fundamental scaling relation for non-interacting spiral galaxies.
An absolute
blue magnitude versus H~{\small{I}} line width
version of the Tully-Fisher relationship is shown in Figure 7.
We have compared our local LCBGs with the Tully-Fisher 
relation found for $\sim$4500 normal galaxies
within $\sim$40 Mpc
\citep{tul00}.    

As seen in
Figure 7, the location of local LCBGs is consistent with
the \cite{tul00} relation, although with a large scatter.
\cite{tul00} have a 1 $\sigma$ scatter of 0.3 magnitudes in M$_B$, while
the local LCBGs have a 1 $\sigma$ scatter of 0.9 magnitudes in M$_B$. 
Approximately half the galaxies (without other sources in the beam)
lie to the left of the Tully-Fisher relationship, indicating
they have lower masses than expected from their luminosities, although this
result is not statistically significant.  This is
consistent with the distribution of LCBG and Hubble type galaxy
mass-to-light ratios in Figure 6.
Seven of the galaxies lying to the right of the 
Tully-Fisher
relationship have other galaxies within the GBT beam,
possibly leading to overestimations of the line widths. 

Note that in our comparison of magnitudes, masses, and mass-to-light
ratios, as in all other comparisons in this paper, we have adjusted all 
values to a Hubble constant of 
H$_0$~=~70~km~s$^{-1}$~Mpc$^{-1}$

\section{Discussion: Comparison with Local Galaxy Types}
We have observed 20 local LCBGs chosen with the same
selection criteria as 
those LCBGs common at higher redshifts. 
LCBGs cannot remain LCBGs for a long period of time:
The number density of LCBGs decreases by at least a factor of ten from z $>$ 0.5
to today.  That is, while LCBGs were common in the past, there are
very few today. Out of approximately a million
nearby galaxies observed by the SDSS, only about a hundred
are LCBGs.  (See Castander et al. 2004 for a discussion of the local space
density of LCBGs.) 
Therefore, LCBGs must
evolve into
some other galaxy type.
 
From studies of intermediate redshift LCBGs,
\cite{koo94} and \cite{guz96} proposed that some may be progenitors
of local
dwarf elliptical galaxies.  Alternatively, 
\cite{phi97} and \cite{ham01} proposed that some may
be disk galaxies in the process of forming a bulge to
become present-day
L$^\star$ spiral galaxies.  
Two pieces of information are crucial to determining if
these possibilities are likely.  It is necessary
to know the
dynamical
masses and the duration of the
starbursts.
We have measured the dynamical masses of our local sample of
LCBGs
using H~{\small{I}}.  The length of the
starburst
will not be estimated until we discuss our molecular
gas survey in Paper II \citep{gar04}.  However, we can use our measurements
of line width and radius to identify local galaxies comparable to LCBGs, independent of
the evolutionary stage of their stellar populations.

In Figure 8 we plot the effective or half-light radius, R$_e$,
versus the velocity dispersion, $\sigma$, 
for our sample of local LCBGs, intermediate redshift (0.4 $\lesssim$ z $\lesssim$ 1)
LCBGs \citep{guz97, phi97},
and local samples of elliptical,
spiral, Magellanic spiral, irregular, and dwarf elliptical galaxies 
(Guzm\'an et al. 1996; HyperLeda).
We measured the $\sigma$ of our sample of galaxies from the H~{\small{I}}
spectra by measuring the moments of the spectra.  The measurements
were made in the same way as we described in $\S$3.2 for measuring
$\int~{S~dv}$, the zeroth moment.  We find similar results if
we simply scale W$_{20}$ to $\sigma$ by assuming the spectra are
Gaussian, that is, dividing W$_{20}$ by 3.6.  The
values of $\sigma$ for the other galaxy samples are from optical
line width measurements, except the measurements for Magellanic spirals
are from 21~cm H~{\small{I}} line widths.
Optical line widths may be smaller than H~{\small{I}}
line widths.  For example, \cite{pis01} studied a sample of nearby
blue compact galaxies and found the ionized gas emission line widths  
to be systematically smaller than the neutral hydrogen emission line 
widths.  On average, the ratio of W$_{20}$ measured from H {\small{II}} to
W$_{20}$ measured from H~{\small{I}} was 0.66 for their sample of 11 blue
compact galaxies.  However, the galaxies in \cite{pis01}'s sample
tend to be smaller both in effective radius and H~{\small{I}} line
width than our sample of local LCBGs, suggesting that the 
difference between the optical and H~{\small{I}} line widths may be
smaller as well \citep{pis01, mar98}.

The R$_e$ versus $\sigma$ plot allows us to 
compare the dynamical mass properties of our sample of LCBGs with 
the other galaxy types.  These properties are expected to remain 
constant despite luminosity 
evolution.  We find that while some local LCBGs
have $\sigma$ consistent with local spiral galaxies,
they tend to be too small in R$_e$.  However, local LCBGs
are consistent with the smaller spiral galaxies, Magellanic spirals.
They are 
inconsistent with elliptical galaxies, but are consistent
with the most massive dwarf elliptical  and irregular galaxies.  There is much 
confusion by what various authors mean when discussing dwarf
elliptical galaxies (sometimes called spheroidal galaxies).  We are
referring to galaxies such as NGC~205 and \emph{not} the less massive and
less luminous galaxies like Draco and Carina.
Finally,
when we compare our local sample of LCBGs with those LCBGs observed at
intermediate redshifts, we find that they occupy the same region
of R$_e$~$-$~$\sigma$ space, suggesting we are indeed examining
a similar mass range in both samples of galaxies.

We also compared local LCBGs to other galaxy types in W$_{20}$~sin$^{-1}$(\emph{i})
versus R$_{25}$ space, where W$_{20}$ was measured from H~{\small{I}}
emission for all samples.  This allowed us to investigate if the wavelength
used to measure the line width, or correcting the line width
for inclination, was having
an effect on our interpretation of which galaxy types most resemble LCBGs.
We could only compare local LCBGs, spirals, Magellanic spirals, and irregular
galaxies, as H~{\small{I}} observations of
intermediate redshift LCBGs and dwarf ellipticals 
are rare or non-existent.
Our findings were entirely consistent with
the interpretation from Figure 8.

Local LCBGs are therefore consistent with the dynamical mass
properties of the most massive dwarf ellipticals and irregulars, and
lower mass or Magellanic spirals.  These classes vary in color and
magnitude.  Knowledge of the amount of molecular gas, time
scale of starburst
and fading, along with the ability of the galaxy to retain
its interstellar medium, will
allow us to discriminate between these remaining
possibilities.  We will begin
this work in Paper II \citep{gar04}.  Given the diverse nature of LCBGs,
it is
likely that multiple scenarios apply, each to a different
subset of 
LCBGs.

\section{Conclusions}
We have performed a single dish 21~cm H~{\small{I}} survey
of
20 local LCBGs chosen to be local analogs to the numerous
LCBGs studied at intermediate redshifts (0.4~$\lesssim$~z~$\lesssim$ 0.7).  
Our findings have verified results from intermediate
redshift LCBG studies.
We have found that local LCBGs are a morphologically heterogeneous mixture
of galaxies.  They are typically gas-rich, with median values of
M$_{HI}$~=~5$\times$10$^9$~M$_\odot$ 
and M$_{HI}$~L$_B^{-1}$~=~0.4~M$_\odot$~L$_\odot^{-1}$.  
Approximately half 
have mass-to-light ratios 
approximately ten times smaller than local galaxies of all Hubble
types at similar
luminosities, confirming
that these are indeed small galaxies
undergoing vigorous bursts of star formation.  This proportion
is likely an underestimate, as nearly half our sample of galaxies
may have dynamical mass overestimates.

By comparing line widths and radii with local galaxy populations, 
we find that local LCBGs are consistent with Magellanic spirals, and the more massive
irregulars and 
dwarf ellipticals.  Measurements
of the length of starburst, amount of fading, and ability of
these galaxies
to retain their interstellar media will help to constrain the evolutionary
possibilities of this galaxy class.
We begin to address these issues in Paper II \citep{gar04}, where we
present the results of
a molecular gas survey of these same local LCBGs.

\bigskip

\noindent
\emph{Acknowledgments} 
We thank the referee for helpful comments which improved the quality
of this paper. 
We thank Rick Fisher for providing the H~{\small{I}} spectrum
of SDSSJ0934+0014.  We also thank the operators and staff at the 
GBT for their help with the observing and reduction, and their
hospitality.
Support for this work was provided
by the NSF through award GSSP02-0001
from the NRAO.  D.~J.~P. acknowledges generous support from an NSF MPS Distinguished
International Postdoctoral Research Fellowship, NSF Grant AST0104439.
R.~G. acknowledges funding from NASA grant LTSA NAG5-11635. 
Funding for the creation and distribution of the SDSS Archive has 
been provided by the Alfred P. Sloan Foundation, the Participating 
Institutions, the National Aeronautics and Space Administration, 
the National Science Foundation, the U.S. Department of Energy, 
the Japanese Monbukagakusho, and the Max Planck Society. The SDSS Web 
site is http://www.sdss.org/.  
We have made extensive use of HyperLeda 
(http://www-obs.univ-lyon1.fr/hypercat/) and the NASA/IPAC
Extragalactic 
Database (NED) 
which is operated by the Jet Propulsion Laboratory, 
California Institute of Technology, 
under contract with the National Aeronautics and Space
Administration
(http://nedwww.ipac.caltech.edu/).  
The Digitized Sky Surveys were produced at the Space Telescope Science 
Institute under U.S. Government grant NAG W-2166. The images of these surveys 
are based on photographic data obtained using the Oschin Schmidt Telescope on 
Palomar Mountain and the UK Schmidt Telescope.

\begin{deluxetable}{llcccclcc}
\rotate
\tabletypesize{\scriptsize}
\tablecaption{Local LCBG Properties\label{tbl-1}}
\tablewidth{0pt}
\tablehead{
\colhead{Source} 			& 
\colhead{Alternate Names} 			&
\multicolumn{1}{c}{D$_{OPT}$} 			& 
\colhead{M$_B$} 			& 
\colhead{B$-$V} 			&
\colhead{SBe} 				&
\colhead{NED Type} 			& 
\colhead{HyperLeda Type} 			& 
\colhead{Other Sources}			\\
\colhead{}				&
\colhead{}				&
\colhead{(Mpc)}				&
\colhead{}				&
\colhead{}				&
\colhead{(B-mag arc sec$^{-2}$)}		&
\colhead{}				&
\colhead{}				&
\colhead{in Beam?}				
}
\startdata
Mrk 297 & NGC 6052, NGC 6064   & 67 &  $-$21.0 &  0.4 &  20.6 &
~ & Sc  & y \\
Mrk 314 & NGC 7468  & 30 &  $-$18.5 &  0.4 &  20.2 &
E3, pec (polar ring?) & E & n \\
Mrk 325 & NGC 7673, Mrk 325            & 49 &  $-$20.0 &  0.4 & 
20.0 & SAc?, pec, H II starburst & Sc    & y \\
Mrk 538 & NGC 7714   		    & 40 &  $-$20.1 &  0.4 &  20.2 &
SB(s)b, pec, H II liner   & SBb & y \\
SDSS J011932.95+145219.0 & NGC 469             & 59 & 
$-$18.9 &  0.4 &  20.3 & &    &  y \\
SDSS J021808.75$-$075718.0 &    		    & 69 &  $-$18.8 &  0.5
&  20.2 & &    & n \\
SDSS J022211.96$-$083036.2 &                    & 67 & 
$-$18.6 &  0.4 &  20.1 & &    & n \\
SDSS J072849.75+353255.2 &    &           56 &  $-$18.9 & 
0.4 &  20.3 & S? & Sbc   & n \\
SDSS J083431.70+013957.9 &    &           59 &  $-$19.1 & 
0.6 &  20.6 & SB(s)b & SBb   & y \\
SDSS J090433.53+513651.1 & Mrk 101   &           68 & 
$-$19.7 &  0.6 &  20.3 &  S & Sc    & n \\
SDSS J091139.74+463823.0 & Mrk 102   &           61 & 
$-$19.3 &  0.5 &  19.1 & S?  &  & n \\
SDSS J093410.52+001430.2 & Mrk 1233   &           70 & 
$-$19.8 &  0.3 &  19.9 &  Sb & Sbc  & y \\
SDSS J093635.36+010659.8 &    &           71 &  $-$19.1 & 
0.6 &  21.0 &   &  & y \\
SDSS J094302.60$-$021508.9 &    &           68 &  $-$19.3 & 
0.5 &  20.4 & S0  & S0  & n \\
SDSS J111836.35+631650.4 & Mrk 165   &           46 & 
$-$18.6 &  0.4 &  19.4 & compact starburst &     & n \\
SDSS J123440.89+031925.1 & NGC 4538   &           67 & 
$-$19.2 &  0.6 &  21.0 & S pec  & Sbc   & n \\
SDSS J131949.93+520341.1 &    &           67 &  $-$18.7 & 
0.2 &  20.1 &  &    & y \\
SDSS J140203.52+095545.6 & NGC 5414, Mrk 800   &           61
&  $-$19.7 &  0.5 &  19.7 & pec  &  & y \\
SDSS J150748.33+551108.6 &    &           48 &  $-$18.9 & 
0.4 &  20.8 & S & Sbc   & n \\
SDSS J231736.39+140004.3 & NGC 7580, Mrk 318   &           63
&  $-$19.3 &  0.6 &  20.4 &  S? & Sbc    & n \\
\enddata
\end{deluxetable}

\begin{deluxetable}{lllll}
\tabletypesize{\footnotesize}
\tablecaption{Local LCBG H~{\small{I}} Line Measurements
\label{tbl-2}}
\tablewidth{0pt}
\tablehead{
\colhead{Source} 			& 
\colhead{V$_{BARY}$} 			&
\multicolumn{1}{c}{D} 			& 
\colhead{W$_{20}$} 			& 
\colhead{$\int$ S dv}			\\
\colhead{}				&
\colhead{(km s$^{-1}$)}				&
\colhead{(Mpc)}				&
\colhead{(km s$^{-1}$)}			&	
\colhead{(Jy km s$^{-1}$)}			
}
\startdata
Mrk 297	&	4739	$\pm$	17	&	68	$\pm$	0.2	&	362	$\pm$	17	&	6.5	$\pm$	0.08	\\
Mrk 314	&	2081	$\pm$	17	&	30	$\pm$	0.2	&	188	$\pm$	17	&	12	$\pm$	0.1	\\
Mrk 325	&	3427	$\pm$	17	&	49	$\pm$	0.2	&	202	$\pm$	17	&	11	$\pm$	0.2	\\
Mrk 538	&	2798	$\pm$	17	&	40	$\pm$	0.2	&	240	$\pm$	17	&	20	$\pm$	0.2	\\
SDSSJ0119+1452	&	4098	$\pm$	17	&	59	$\pm$	0.2	&	266	$\pm$	17	&	2.4	$\pm$	0.1	\\
SDSSJ0218$-$0757	&		&		&				&	0.41	$\pm$	0.07	\\
SDSSJ0222$-$0830	&		&		&				&	0.61	$\pm$	0.09	\\
SDSSJ0728+3532	&	3953	$\pm$	17	&	56	$\pm$	0.2	&	216	$\pm$	17	&	7.9	$\pm$	0.1	\\
SDSSJ0834+0139 	&	4215	$\pm$	160	&	60	$\pm$	2	&	329	$\pm$	160	&	6.9	$\pm$	0.2	\\
SDSSJ0904+5136	&	4782	$\pm$	103	&	68	$\pm$	2	&	204	$\pm$	103	&	4.5	$\pm$	0.2	\\
SDSSJ0911+4636	&	4281	$\pm$	17	&	61	$\pm$	0.2	&	151	$\pm$	17	&	2.0	$\pm$	0.1	\\
SDSSJ0934+0014	&	4860	$\pm$	17	&	69	$\pm$	0.2	&	332	$\pm$	17	&	4.8	$\pm$	0.2	\\
SDSSJ0936+0106	&	4920	$\pm$	17	&	70	$\pm$	0.2	&	255	$\pm$	17	&	3.3	$\pm$	0.09	\\
SDSSJ0943$-$0215	&	4823	$\pm$	17	&	69	$\pm$	0.2	&	230	$\pm$	17	&	2.6	$\pm$	0.06	\\
SDSSJ1118+6316	&	3218	$\pm$	17	&	46	$\pm$	0.2	&	152	$\pm$	17	&	2.2	$\pm$	0.07	\\
SDSSJ1234+0319	&	4685	$\pm$	17	&	67	$\pm$	0.2	&	269	$\pm$	17	&	4.4	$\pm$	0.1	\\
SDSSJ1319+5203	&	4619	$\pm$	17	&	66	$\pm$	0.2	&	202	$\pm$	17	&	7.7	$\pm$	0.09	\\
SDSSJ1402+0955	&	4267	$\pm$	17	&	61	$\pm$	0.2	&	343	$\pm$	17	&	6.8	$\pm$	0.2	\\
SDSSJ1507+5511	&	3373	$\pm$	17	&	48	$\pm$	0.2	&	126	$\pm$	17	&	3.7	$\pm$	0.1	\\
SDSSJ2317+1400	&	4413	$\pm$	17	&	63	$\pm$	0.2	&	254	$\pm$	17	&	6.2	$\pm$	0.1	\\
\enddata
\tablenotetext{a} { Marginal detection; no line width measurements.}
\end{deluxetable}

\begin{deluxetable}{llccccccc}
\rotate
\tabletypesize{\scriptsize}
\tablecaption{Derived Parameters for Local
LCBGs\label{tbl-3}}
\tablewidth{0pt}
\tablehead{
\colhead{Source} 			&   
\multicolumn{1}{c}{M$_{HI}$} 		&
\colhead{Inclination} 			& 
\colhead{W$_R^i$}			&
\colhead{V$_{ROT}$} 			& 
\colhead{R$_{25}$(B)} 			&
\colhead{M$_{DYN}$ (R $<$ R$_{25}$(B))} &
\colhead{R$_e(r)$} 			&
\colhead{M$_{DYN}$ (R $<$ R$_e$(r))} 	\\
\colhead{} 				&
\colhead{(10$^9$ M$_\odot$)} 		&
\colhead{($^\circ$)} 			&
\colhead{(km s$^{-1}$)}			&
\colhead{ (km s$^{-1}$)} 		&
\colhead{(kpc)} 			&
\colhead{(10$^{10}$ M$_\odot$)} 	&
\colhead{(kpc)} 			&
\colhead{(10$^{10}$ M$_\odot$)}
}
\startdata
Mrk 297 &  7.0 $\pm$ 0.1 &  42\tablenotemark{a} & 486 &  243 &  8.0 &  11 &  2.0\tablenotemark{c} &  2.7 \\
Mrk 314 &  2.5 $\pm$ 0.05 &  65\tablenotemark{a} &  170 & 85 &  3.8 &  0.63 &  0.94\tablenotemark{c} &  0.16 \\
Mrk 325 &  6.3 $\pm$ 0.1 &  43\tablenotemark{a} &  242 & 121 &  9.6 &  3.3 &  2.4\tablenotemark{c} &  0.83 \\
Mrk 538 &  7.6 $\pm$ 0.1 &  50\tablenotemark{a} &  264 & 132 &  11 &  4.5 &  2.8\tablenotemark{c} &  1.1 \\
SDSSJ0119+1452 &  2.0 $\pm$ 0.1 &  84 & 230 & 115 &  5.6 &  1.7 &  1.6 & 
0.48 \\
SDSSJ0218$-$0757 &  0.47 $\pm$ 0.08 &  52 & ~ & ~ &  4.9\tablenotemark{b} & 
~ &  1.3 &  ~ \\
SDSSJ0222$-$0830 &  0.65 $\pm$ 0.1 &  48 & ~ & ~ &  4.2\tablenotemark{b} & 
~ &  1.2 &  ~ \\
SDSSJ0728+3532 &  6.0 $\pm$ 0.1 &  34 & 321 & 160 &  5.4 &  3.3 &  1.2 & 
0.74 \\
SDSSJ0834+0139 &  5.9 $\pm$ 0.5 &  83 & 293 & 147 &  6.5 &  3.3 &  1.6 & 
0.82 \\
SDSSJ0904+5136 &  4.9 $\pm$ 0.3 &  34 & 301 & 150 &  7.4 &  3.9 &  2.0 & 
1.0 \\
SDSSJ0911+4636 &  1.7 $\pm$ 0.1 &  33 &  221 & 111 &  5.6 &  1.6 &  1.2 & 
0.35 \\
SDSSJ0934+0014 &  5.4 $\pm$ 0.2 &  51 &  378 & 189 &  6.8 &  5.7 &  1.9 & 
1.5 \\
SDSSJ0936+0106 &  3.8 $\pm$ 0.1 &  48 &  293 & 146 &  7.1 &  3.5 &  2.0 & 
1.0 \\
SDSSJ0943$-$0215 &  2.9 $\pm$ 0.07 &  84 &  194 & 97 &  4.9 &  1.1 &  1.8 & 
0.39 \\
SDSSJ1118+6316 &  1.1 $\pm$ 0.04 &  82 &  123 & 61 &  3.1\tablenotemark{b} &  0.27 &  0.91 & 
0.08 \\
SDSSJ1234+0319 &  4.7 $\pm$ 0.1 &  52 &  293 & 147 &  6.7 &  3.4 &  2.1 & 
1.1 \\
SDSSJ1319+5203 &  7.9 $\pm$ 0.1 &  44 &  239 & 120 &  5.3 &  1.8 &  1.2 & 
0.41 \\
SDSSJ1402+0955 &  6.0 $\pm$ 0.2 &  55 &  372 & 186 &  8.7 &  7.0 &  1.6 & 
1.3 \\
SDSSJ1507+5511 &  2.0 $\pm$ 0.07 &  44 &  144 & 72 &  7.9 &  0.95 &  1.8 & 
0.21 \\
SDSSJ2317+1400 &  5.8 $\pm$ 0.1 &  31 &  420 & 210 &  7.3 &  7.5 &  1.7 & 
1.8 \\
\enddata
\tablenotetext{a} { from HyperLeda}
\tablenotetext{b} { estimated as 4 $\times$ R$_e$(B)}
\tablenotetext{c} { estimated as 0.25 $\times$ R$_{25}$}
\end{deluxetable}

\begin{deluxetable}{lccccc}
\tabletypesize{\scriptsize}
\tablecaption{Additional Derived Parameters for Local LCBGs
\label{tbl-4}}
\tablewidth{0pt}
\tablehead{
\colhead{Source} 				&   
\multicolumn{1}{c}{M$_{HI}$ M$_{DYN}^{-1}$(R $<$ R$_{HI}$)}  	&
\colhead{L$_B$} 				& 
\colhead{M$_{HI}$ L$_B^{-1}$} 			& 
\colhead{M$_{DYN}$(R $<$ R$_{25}$) L$_B^{-1}$} 	&
\colhead{M$_{DYN}$(R $<$ R$_{e}$) L$_B^{-1}$}  	\\
\colhead{} 					&
\colhead{} 					&
\colhead{(10$^9$ L$_\odot$)} 				&
\colhead{(M$_\odot$ L$_\odot^{-1}$)}			&
\colhead{(M$_\odot$ L$_\odot^{-1}$)} 		&
\colhead{(M$_\odot$ L$_\odot^{-1}$)} 		
}
\startdata
Mrk 297 &  0.03 &  39 &  0.2 &  3 &  0.7 \\
Mrk 314 &  0.2 &  3.9 &  0.7 &  2 &  0.4 \\
Mrk 325 &  0.1 &  16 &  0.4 &  2 &  0.5 \\
Mrk 538 &  0.08 &  17 &  0.5 &  3 &  0.7 \\
SDSSJ0119+1452 &  0.06 &  5.6 &  0.4 &  3 &  0.9 \\
SDSSJ0218-0757 &  ~ &  5.2 &  0.09 &  ~ &  ~ \\
SDSSJ0222-0830 &  ~ &  4.3 &  0.2 &  ~ &  ~ \\
SDSSJ0728+3532 &  0.09 &  5.6 &  1 &  6 &  1 \\
SDSSJ0834+0139 &  0.1 &  6.8 &  0.9 &  5 &  1 \\
SDSSJ0904+5136 &  0.07 &  12 &  0.4 &  3 &  0.9 \\
SDSSJ0911+4636 &  0.05 &  8.2 &  0.2 &  2 &  0.4 \\
SDSSJ0934+0014 &  0.05 &  13 &  0.4 &  4 &  1 \\
SDSSJ0936+0106 &  0.06 &  6.8 &  0.6 &  5 &  2 \\
SDSSJ0943-0215 &  0.1 &  8.2 &  0.4 &  1 &  0.5 \\
SDSSJ1118+6316 &  0.2 &  4.3 &  0.3 &  0.6 &  0.2 \\
SDSSJ1234+0319 &  0.07 &  7.4 &  0.6 &  5 &  1 \\
SDSSJ1319+5203 &  0.2 &  4.7 &  2 &  4 &  0.9 \\
SDSSJ1402+0955 &  0.04 &  12 &  0.5 &  6 &  1 \\
SDSSJ1507+5511 &  0.1 &  5.6 &  0.4 &  2 &  0.4 \\
SDSSJ2317+1400 &  0.04 &  8.2 &  0.7 &  9 &  2 \\
\enddata
\end{deluxetable}

\newpage

\begin{figure}
\plotone{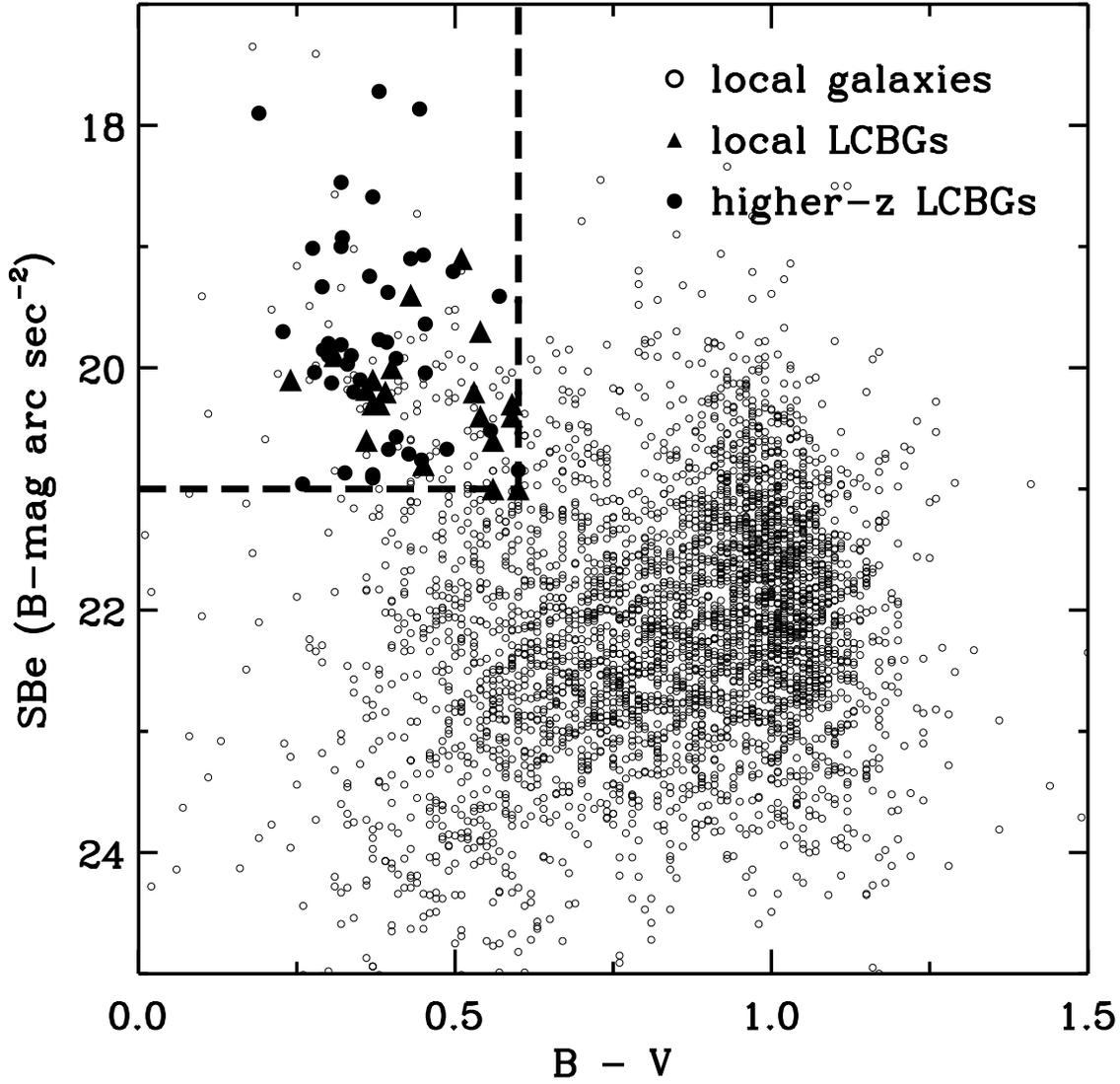}
\caption{B$-$V color versus SBe, average surface brightness
within
the effective radius (B-mag arc sec$^{-2}$),
for three samples of galaxies.  Open circles show
the properties of $\sim$4000 galaxies in the local Universe
(HyperLeda); filled circles represent
our local sample of galaxies which 
were selected with the
same
criteria used to define
intermediate redshift LCBGs;
triangles show a small sample of LCBGs at
higher
redshifts (0.4 $\lesssim$ z $\lesssim$ 1) observed by 
\cite{koo94}, Guzm\'an et al. (1997, 1998), and
\cite{phi97}.
The dotted lines 
roughly demark 
the
color and surface brightness criteria from \cite{jan03}
which separate LCBGs
from normal
galaxies.
Compared to other galaxy types which dominate today's galaxy
population,
LCBGs are relatively rare in the nearby Universe.  
\label{fig1}}
\end{figure}

\begin{figure}
\centerline{
\includegraphics {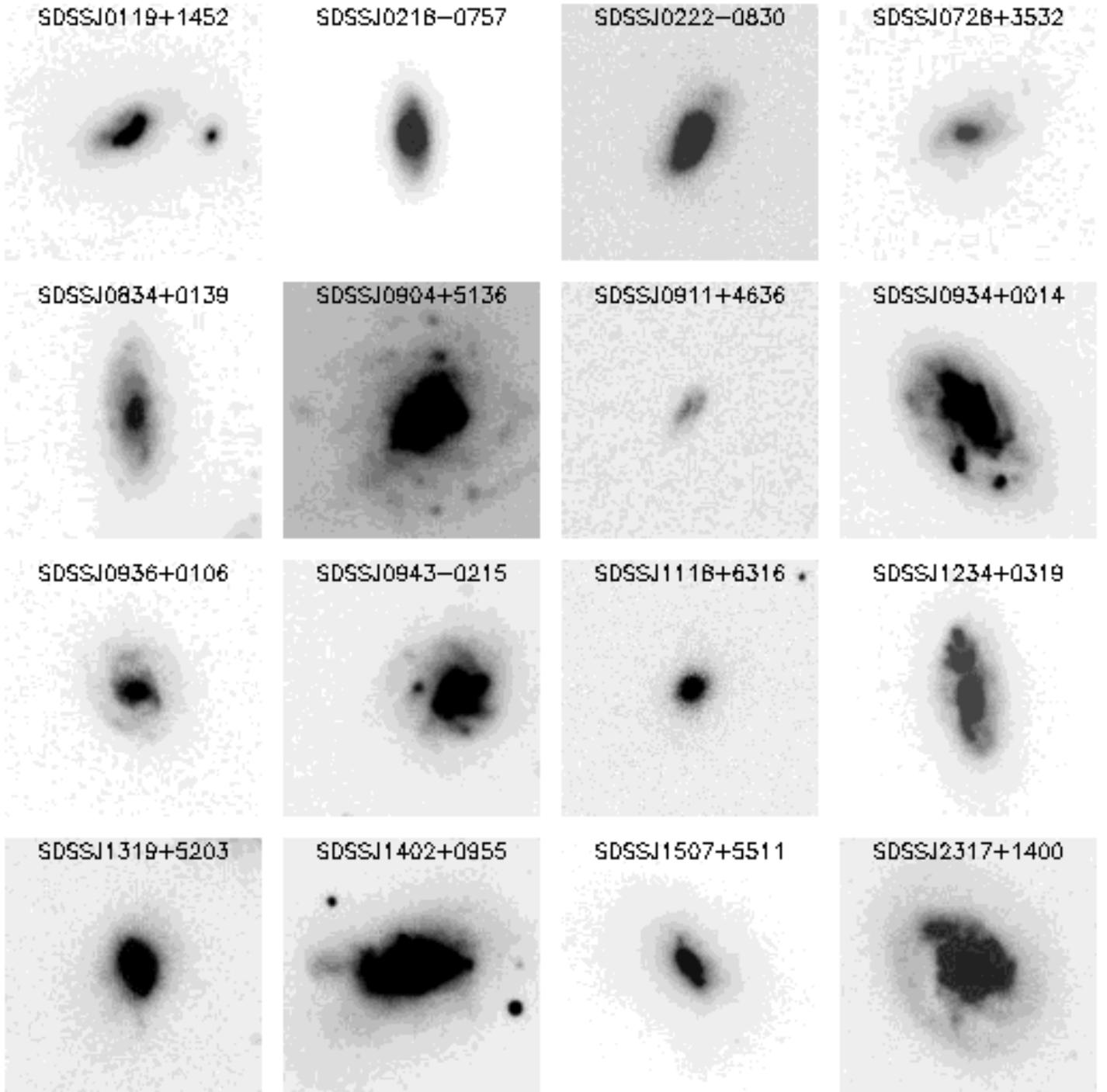}
}
\caption{SDSS g-band (4770~\AA) images of the 16 SDSS local LCBGs. 
Each
image is 15 kpc in diameter.  LCBGs are clearly
morphologically heterogeneous.
\label{fig2}}
\end{figure}

\begin{figure}
\centerline{
\includegraphics {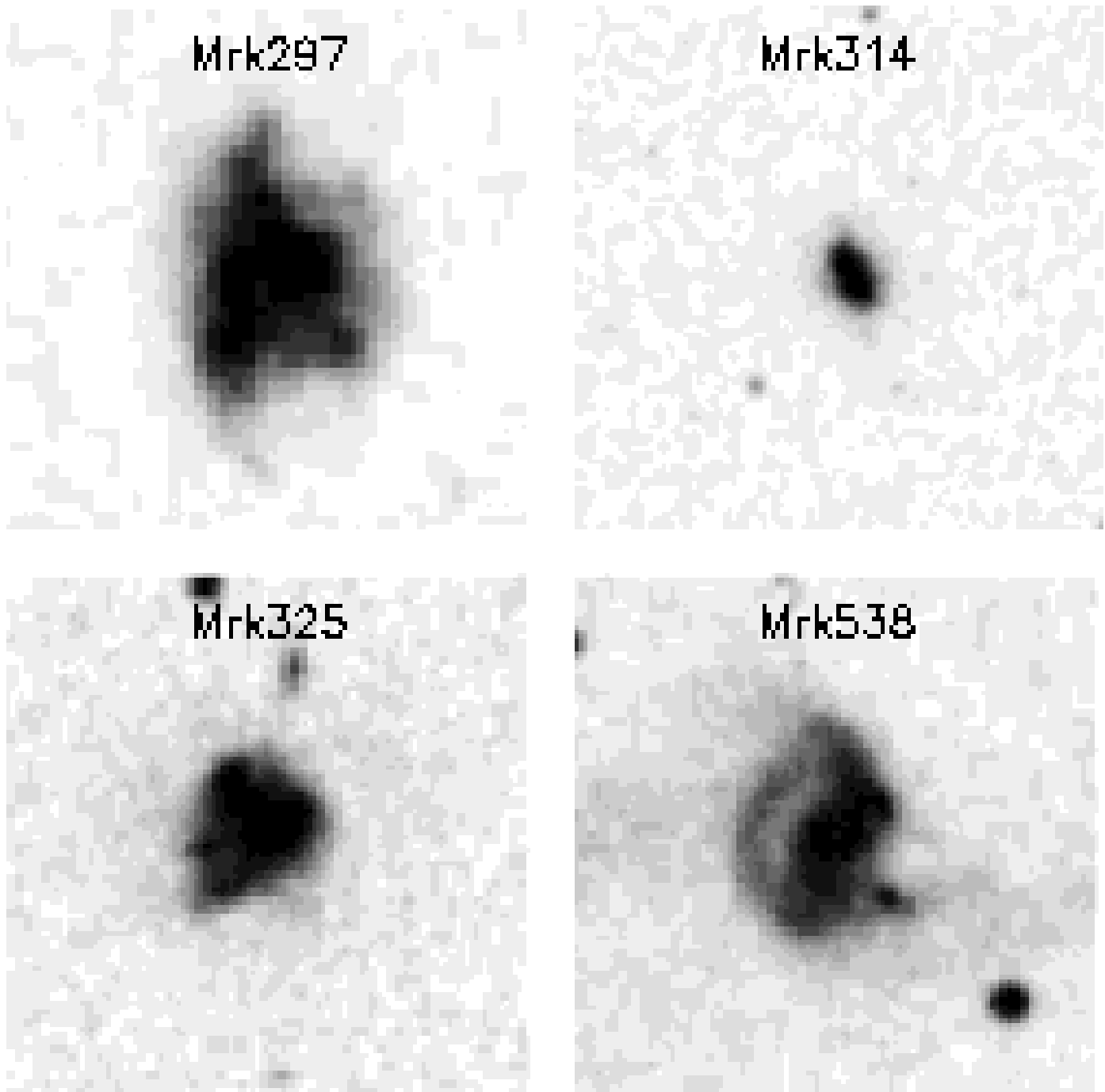}
}
\caption{Digitized Sky Survey images of the four non-SDSS
local LCBGs.
Each image is 35 kpc in diameter.
\label{fig3}}
\end{figure}

\begin{figure}
\centerline{
\includegraphics {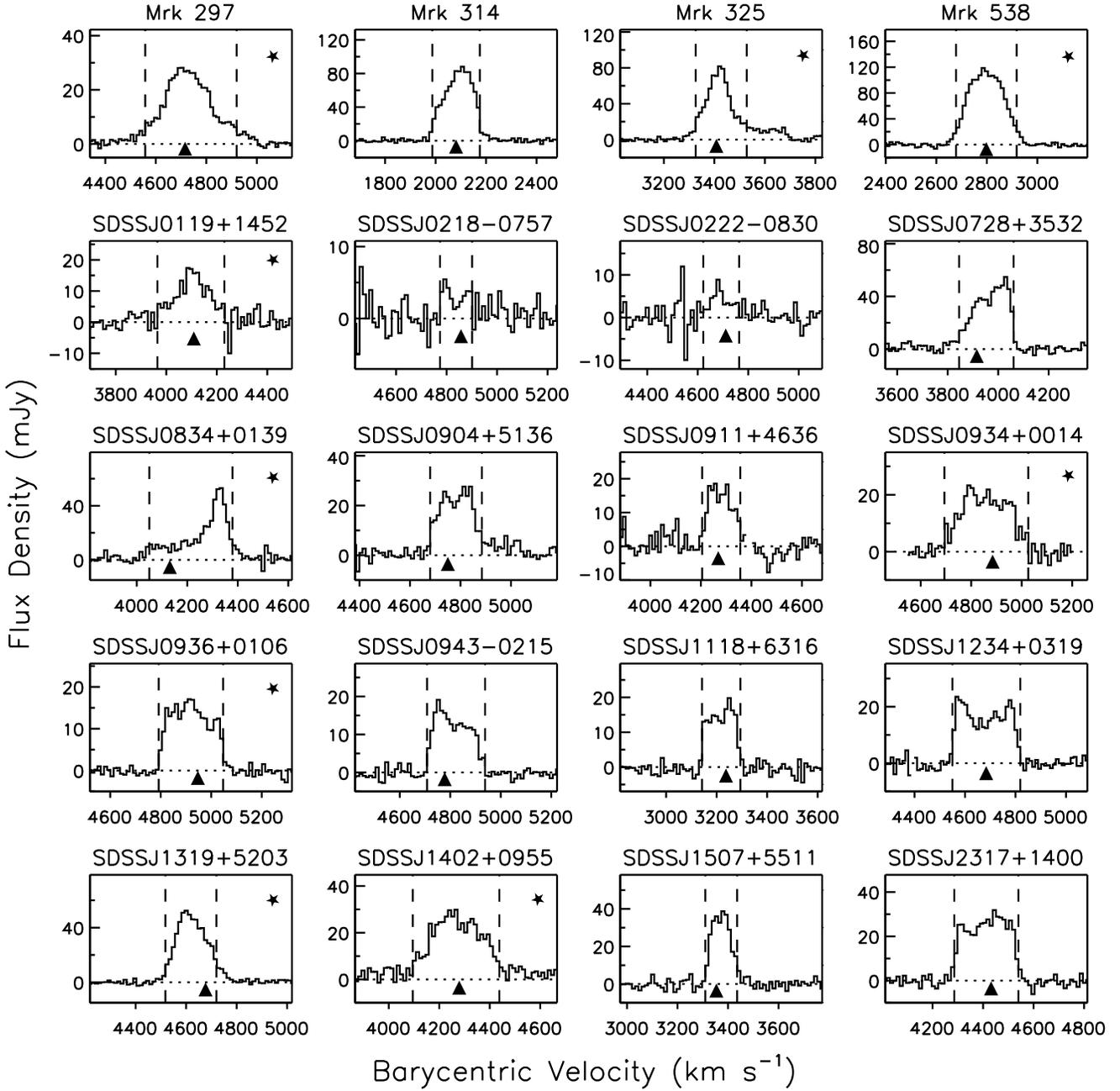}
}
\caption{
GBT H~{\small{I}} spectra of the sample of nearby
LCBGs. The vertical scales are flux densities, in mJy; the 
horizontal scales are barycentric velocities, in km~s$^{-1}$.
The spectra have been smoothed to a resolution of 
$\sim$12~km~s$^{-1}$ and only
the central 800~km~s$^{-1}$ are shown. 
The dashed lines indicate the 20\% crossings used to measure
the line width at 20\%.
The triangles indicate
the recessional velocities calculated from SDSS redshifts for
the SDSS galaxies; velocities from NED redshifts are shown for the 
non-SDSS galaxies.  Those galaxies with other sources at similar velocities
within the GBT beam are indicated by a star in the upper right corner.
\label{fig4}}
\end{figure}

\begin{figure}
\centerline{
\includegraphics{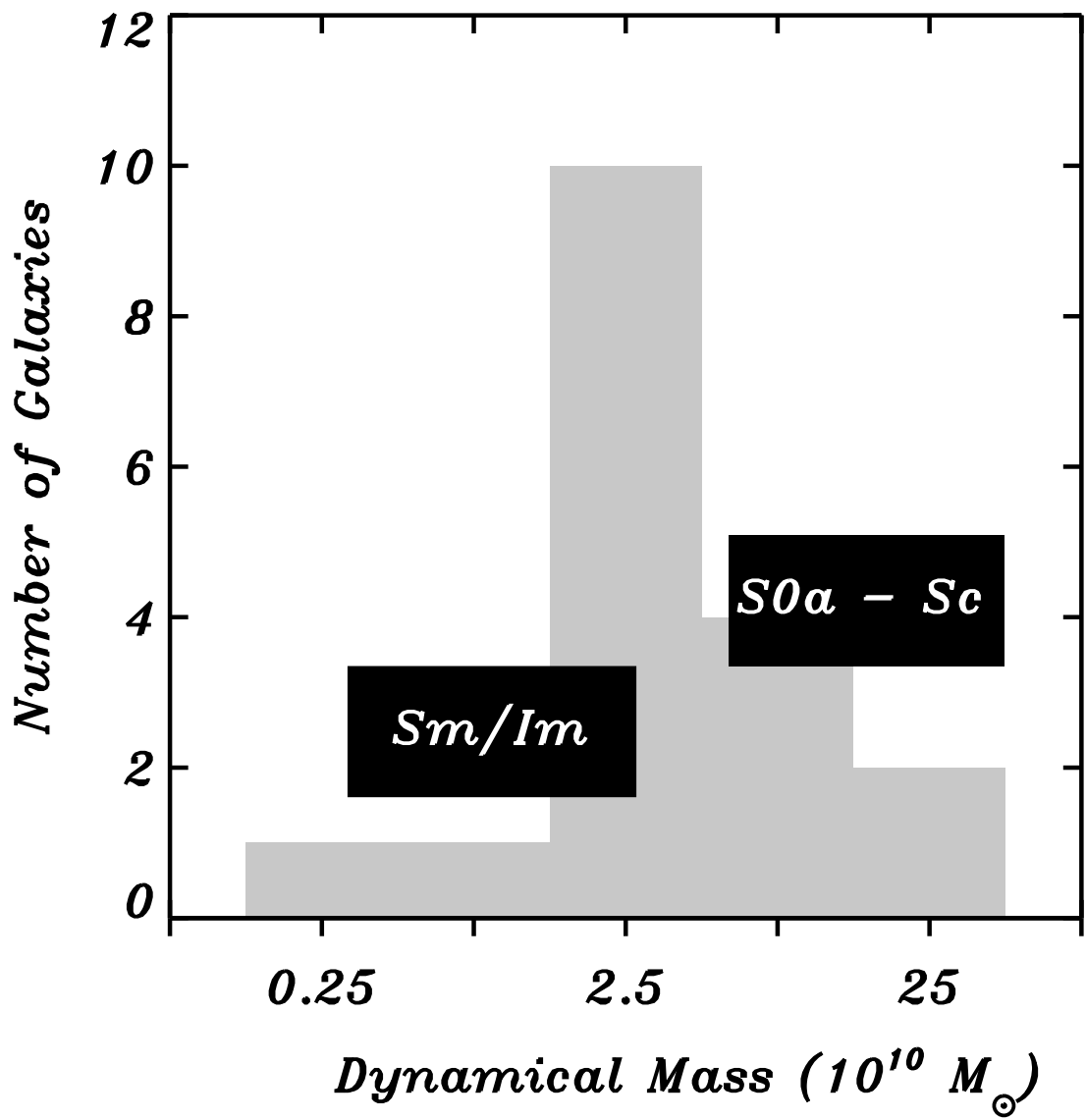}
}
\caption{The distribution of dynamical masses, M$^{NR}_{DYN}$(R~$<~$R$_{25}$),
for local
LCBGs, as measured from 
H~{\small{I}} observations, is shown as a gray histogram. 
For comparison, the range of dynamical masses for
local Hubble type galaxies are indicated with black boxes.
The ``m'' in ``Sm'' and ``Im'' indicates Magellanic
or low-luminosity spirals and irregulars.  
Note that in order
to compare LCBG dynamical masses with \cite{rob94}'s results
for local Hubble types, the dynamical masses plotted here
do not include a line width correction for random motions.
\label{fig5}
}
\end{figure}

\begin{figure}
\centerline{
\includegraphics {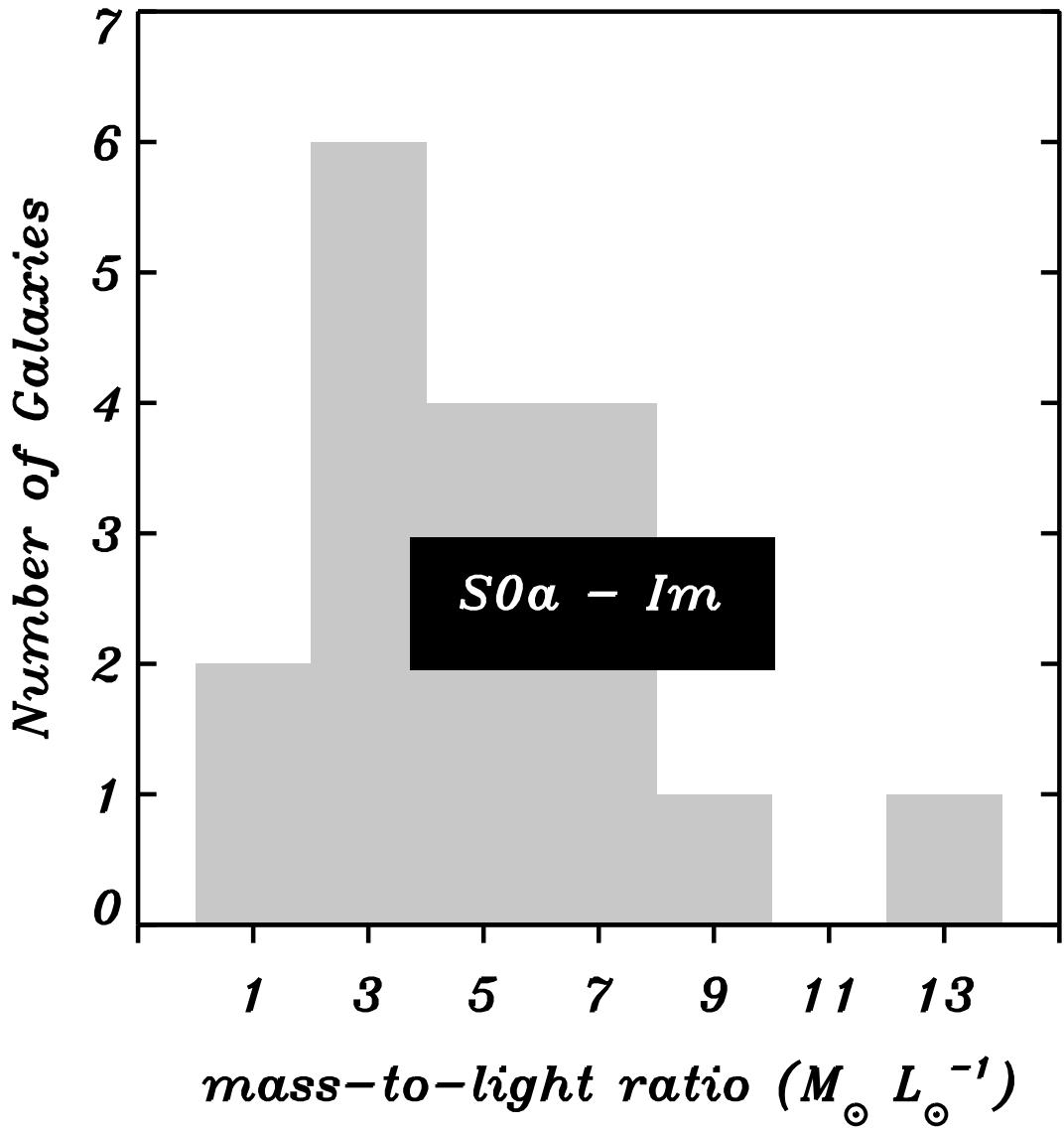}
}
\caption{
The distribution of 
dynamical mass (within R$_{25}$) to L$_B$ ratios for local
LCBGs
is shown as a gray histogram. 
For comparison, the range of mass-to-light
ratios for local Hubble type galaxies (S0a to Im) \citep{rob94} are shown. 
As in Figure 5, for accurate comparisons, these mass-to-light ratios
do not include a line width correction for random motions.
Many LCBGs tend to have mass-to-light ratios
smaller than local normal galaxies, consistent with
findings at intermediate redshifts.  Note that we may have overestimated the
dynamical masses for most of the LCBGs with
higher mass-to-light ratios.
\label{fig6}
}
\end{figure}

\begin{figure}
\centerline{
\includegraphics  {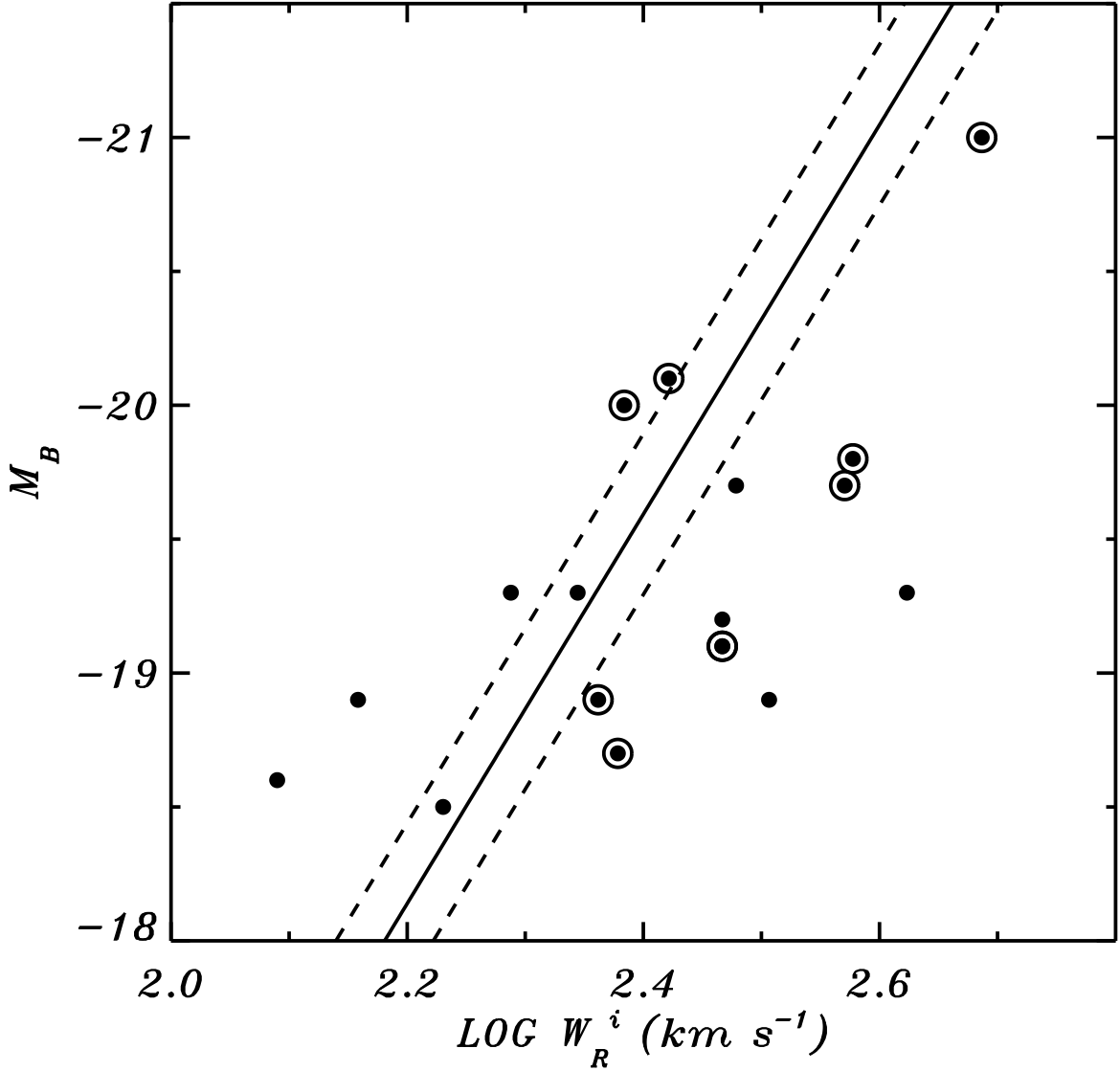}
}
\caption{The Tully-Fisher distribution:  the logarithm of the line width at 20\%
corrected for inclination and random motions (W$_R^i$) versus
absolute blue magnitude (M$_B$).  Local LCBGs are indicated by
filled circles.  Those galaxies with other sources at similar velocities within
the GBT beam are circled; their line widths may be
overestimated.
The solid line indicates
the Tully-Fisher relationship from \cite{tul00}; the dotted lines indicate their
1~$\sigma$ scatter of 0.3~M$_B$.  
The location of our local LCBGs is consistent with the Tully-Fisher
relationship, but with a higher 1~$\sigma$ scatter of 0.9~M$_B$.
\label{fig7}}
\end{figure}

\begin{figure}
\centerline{
\includegraphics {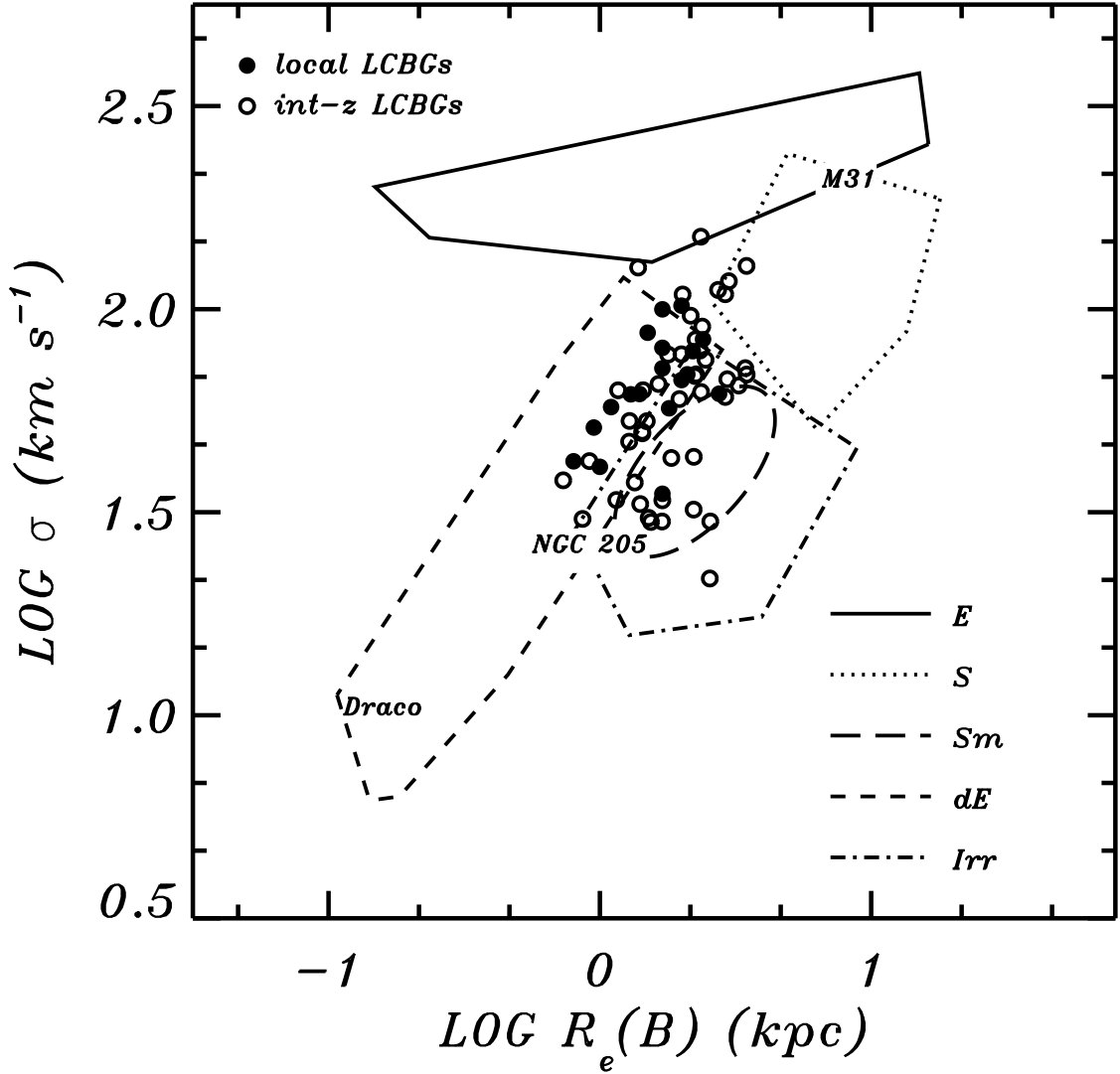}
}
\caption{The effective radius (R$_e$)
versus the the velocity dispersion ($\sigma$) is plotted
for our local LCBGs (filled circles) and intermediate redshift 
(0.4 $\lesssim$ z $\lesssim$ 1)
LCBGs (open circles) \citep{guz97, phi97}.
The regions occupied by the bulk of other local galaxy 
types$-$ellipticals (E), spirals (S), Magellanic spirals (Sm),
dwarf ellipticals (dE), and irregulars (Irr)$-$are indicated 
(Guzm\'an et al. 1996, HyperLeda).  We have also indicated
the approximate locations for some representative galaxies:
Draco, NGC~205 and M~31 (HyperLeda).
Local LCBGs are
consistent with higher mass irregulars and dwarf ellipticals,
and lower mass or Magellanic spirals, 
as well
as intermediate redshift LCBGs.  
\label{fig8}}
\end{figure}

\end{document}